\documentclass[prd,twocolumn,nofootinbib]{revtex4}

\usepackage{amsmath,amssymb,graphicx}

\newcommand{\bm}[1]{\mbox{\boldmath{$#1$}}}

\begin{document}

\preprint{\hepth{0808.3770}}

\title{Properties of the scale factor measure}

\author{Raphael Bousso, Ben Freivogel and I-Sheng Yang\footnote{bousso@lbl.gov, freivogel@berkeley.edu, 
jingking@berkeley.edu}}

\affiliation{Department of Physics and Center for Theoretical
Physics \\
University of California, Berkeley, CA 94720, U.S.A. \\
{\em and}\\
Lawrence Berkeley National Laboratory, Berkeley, CA 94720, U.S.A. }

\begin{abstract}
  We show that in expanding regions, the scale factor measure
  can be reformulated as a local measure: Observations are weighted by
  integrating their physical density along a geodesic that starts in
  the longest-lived metastable vacuum.  This explains why some of its
  properties are similar to those of the causal diamond measure.  In
  particular, both measures are free of Boltzmann brains, subject to
  nearly the same conditions on vacuum stability.  However, the scale
  factor measure assigns a much smaller probability to the observed
  value of the cosmological constant.  The probability decreases
  further, like the inverse sixth power of the primordial density
  contrast, if the latter is allowed to vary.
\end{abstract}

\maketitle

\section{Introduction}

\subsection{The measure problem}

Like the cosmological constant problem~\cite{Wei89,Pol06,TASI07}, the
measure problem arises purely within the regime of validity of
semi-classical gravity.  All that is needed is a long-lived vacuum
state~\cite{GutWei83}, or a sufficiently flat scalar field
potential~\cite{Linde,GonLin87}, with positive vacuum energy.  Under
these conditions, a finite spatial region will inflate eternally,
generating an unbounded four-volume.  All possible events will occur
infinitely many times, and a cutoff is needed to compute
probabilities.

Unlike the cosmological constant problem, the measure problem leaves a
loophole: it is possible that the conditions for eternal inflation do
not actually occur in Nature.  But increasing evidence indicates
otherwise.

Slow-roll inflation is the dominant paradigm explaining the origin of
structure and the large-scale homogeneity and flatness of our
universe.  If we are prepared to believe that the moderately
fine-tuned scalar field potential necessary for driving slow-roll
inflation can arise in Nature, it is hard to imagine that the more
generic feature of a local minimum cannot exist.

Moreover, the observed universe has positive vacuum
energy~\cite{Per98,Rie98,Dun08}.  Unless our vacuum is unstable on a
time scale of order ten billion years---which would require remarkable
tuning---it alone suffices to generate eternal inflation.

Finally, the only extant explanation~\cite{Sak84,Wei87,BP} of the
smallness of the observed dark energy---the cosmological constant
problem---is the existence of a multidimensional landscape of
metastable vacua in string theory.  This is empirical evidence for
string theory, and in particular for eternal inflation.

A theory of everything, if it gives rise to eternal inflation, should
eventually allow us to derive a unique prescription for computing
probability amplitudes from first principles.  Yet our understanding
of string theory, especially in cosmological settings, remains
woefully incomplete.  For now, a top-down solution to the measure
problem seems elusive.  Thus, we advocate following the traditional,
phenomenological approach.

\subsection{Phenomenological approach}

Like any other theory, a compelling measure should be reasonably
well-defined, simple, and general.  In tackling a problem so vast and
unfamiliar, it is natural to seek more specific guiding principles.
For example, lessons from the black hole information paradox motivated
the causal diamond measure~\cite{Bou06,BouFre06a}.  We would be ill
advised, however, to turn our intuition into dogma, insisting
absolutely on {\em theoretical\/} properties that ``any reasonable''
measure ``must'' obey.  Not only would we run the danger of putting in
by hand the answers we wish to get; worse, our wishes may be
misguided.  Surely, for example, any reasonable measure must reward
the volume expansion during slow-roll inflation in any given vacuum?
In fact, this intuitive requirement invites conflict with
experiment~\cite{GarVil05,FelHal05}.  The causal diamond measure 
abandoned volume-weighting, and the requirement seems now to have lost 
its dogmatic status~\cite{DGSV08,Pag08}.

This brings us to the one property of a measure that we can insist on
absolutely: that its {\em experimental\/} predictions not conflict
with observations.  In fact, many innocent-looking measures do
conflict with observation, and violently so.  Some conflicts are
extremely robust, arising almost independently of the properties of
the underlying vacuum landscape.  Thus, they allow us to falsify
measures even while we still have much to learn about the landscape.

For example, the proper time cutoff
~\cite{Lin86a,LinLin94,GarLin94,GarLin94a,GarLin95,Lin06} predicts a
very hot 
universe~\cite{LinLin96,Gut00a,Gut00b,Gut04,Teg05,Lin07,Gut07,BouFre07}
with probability 1 (the ``Boltzmann babies'' or youngness paradox).
Measures that involve counting the number of observers per
baryon~\cite{Efs95,Vil95,Wei96,MarSha97,Vil04,GarSch05,Wei05} predict 
an empty, cold universe~\cite{Pag06,BouFre06b} with probability 1 (the
``Boltzmann brain'' paradox).  Such paradoxes are important tools for
testing and eliminating measure theories.  We say ``paradox'', as if
there was any doubt about the culprit.  In fact, the above paradoxes
represent fatal failures: the measure assigns zero probability to the
observations we actually make.  Such measures are experimentally ruled
out and must be discarded.

Another useful test is the more aptly named
``$Q$-catastrophe''~\cite{GarVil05,FelHal05} alluded to earlier.  In
measures that reward the volume expansion during inflation, such as
Ref.~\cite{GarSch05}, inflationary model parameters generically
receive exponential pressure towards extreme values.  In this case,
anthropic constraints do not suffice to explain the moderate values we
observe for, say, the primordial density perturbation, $Q$.

Other problems are more subtle, or depend on the detailed structure of
the vacuum landscape.  The ``staggering problem'' can arise in the
measure of Ref.~\cite{GarSch05}: for some landscape models, the
measure assigns such unequal probabilities to the cosmological
production of different vacua that most observers live in extreme
environments where their existence is an unlikely
fluctuation~\cite{SchVil06,Sch06,BouYan07,OluSch07,Sch08}.  In
particular, the cosmological constant problem cannot be solved in this
case.

Finally, the value it predicts for the cosmological constant,
$\Lambda$, is an important test of any measure.  This particular
observable is special for two reasons: Its statistical distribution in
the landscape is understood well enough to make detailed quantitative
predictions.  And its value has been measured:
\begin{equation}
\Lambda=1.5\times 10^{-123}~,
\label{eq-lambda}
\end{equation}
in Planck units.

At present, the causal diamond measure is the most successful proposal
phenomenologically.  It avoids Boltzmann babies, Boltzmann brains, and
the staggering problem.  (The absence of Boltzmann brains requires
that all vacua decay faster than they produce such rare
fluctuations~\cite{BouFre06b}---a nontrivial condition on the vacuum
landscape, which, however, may well be satisfied by the string
landscape~\cite{FreLip08}.)  It does not reward volume expansion, thus
avoiding the $Q$-catastrophe and raising the possibility of the
detection of open spatial curvature by future
experiments~\cite{FreKle05}.

The causal diamond measure predicts a value of the cosmological
constant consistent with observations.  That is, the observed value
lies close to the mean and is highly typical in the predicted
probability distribution~\cite{BouHar07}.  This successful prediction
is robust against variations of $Q$~\cite{BouHar07,CliFre07}.

\subsection{Summary and outline}

In this paper, we investigate the scale factor measure~\cite{DGSV08},
which cuts off the universe at the time $\eta$ when a (randomly
chosen) congruence of timelike geodesics has expanded by a volume
factor $\exp(3\eta)$ along each geodesic.  Relative probabilities of
different observational outcomes can be defined by computing the
ratios of the number of times such outcomes occur in the regulated
four-volume, and then taking the regulator $\eta\to\infty$.

Section~\ref{sec-defrate} contains review material and establishes
most of our notation. In Sec.~\ref{sec-def}, we give a detailed
definition of the scale factor measure. A separate prescription is
needed in regions where geodesics contract and intersect with each
other. We review the prescription chosen by De Simone {\em et al.},
who were the first to formulate one carefully. In Sec.~\ref{sec-rate},
we collect other useful definitions and results; in particular, we
review the solution found by Garriga, Schwartz-Perlov, Vilenkin, and
Winitzki~\cite{GarSch05} for the volume distribution of different
vacua, which also applies to the scale factor measure.

In Section~\ref{sec-one}, we determine the shape of the scale factor
cutoff hypersurface in a homogeneous, isotropic, open universe formed
by bubble nucleation inside a parent de~Sitter vacuum.  This is
nontrivial, because the local scale factor along each geodesic, and
the average scale factor in a bubble universe, are two different
objects.

In Section~\ref{sec-OO}, we compute the number of observations below
the cutoff surface by integrating over all bubbles produced prior to
the cutoff.  We begin, in Sec.~\ref{sec-homcount}, by approximating
the bubble interior metric as homogeneous, {\em ignoring local
  gravitational collapse such as occurs during galaxy formation}.  We
find that in this ``no-collapse approximation'', the scale factor
measure leads to a very simple result: The probability of an
observation in some vacuum is proportional to the product of the
probability that a given geodesic in the congruence will enter that
vacuum, times the number density, per physical volume, at which
observations occur.  It is does not depend directly on the time at
which they occur, nor on the amount of volume expansion since the
bubble was produced.

Our result generalizes the results of Ref.~\cite{DGSV08}, encompassing
in a single formula the following important properties of the
(no-collapse!) scale factor measure established by De Simone {\em et
  al.}:\footnote{De Simone {\em et al.}  applied their prescription
  for collapsing regions to the homogeneous collapse of bubbles with
  negative cosmological constant, but implicitly used a no-collapse
  approximation elsewhere, ignoring the turnaround and collapse of
  geodesics in structure-forming regions.  This explains any
  discrepancies between our papers, in particular our less favorable
  conclusion concerning the value of $\Lambda$ predicted by the scale
  factor measure.}
\begin{itemize}
\item[(1)] No youngness paradox, since the physical density of
  Boltzmann babies is negligible.
\item[(2)] No reward for excessive volume expansion during slow-roll
  inflation: Once inflation has made the universe flat enough for
  structure formation, any extra e-foldings will not affect the
  physical density of observers.  Thus, the scale factor measure
  avoids the $Q$-catastrophe.
\item[(3)] The probability distribution for $\Lambda$ is in excellent
  agreement with the observed value, Eq.~(\ref{eq-lambda}): If
  $\Lambda$ had dominated much before the time when observations are
  made (i.e., now), galaxies would now be exponentially dilute and the
  average density of observations would be highly suppressed.  This
  result is stable against variations of the primordial density
  contrast, $Q$.
\end{itemize}

A simple result should have a simple explanation, which we provide in
Sec.~\ref{sec-local}: In the no-collapse approximation (i.e., if all
geodesics in the congruence are always expanding), the scale factor
measure is equivalent to the following prescription\footnote{We are
  using here the approximation of \cite{SchVil06}; we will be more
  precise in the body of the paper.}: {\em Consider a
  single geodesic that starts out in the longest-lived metastable
  vacuum of the landscape, $*$.  Compute the expected number of
  observations of type $\mu$, $d\langle N_\mu\rangle$, occuring along
  a fixed physical volume $dV$ transverse to the geodesic.  The
  relative probability of observations $\mu$ and $\nu$ is $d\langle
  N_\mu\rangle /d\langle N_\nu\rangle$.}  In practice, this can be
accomplished by summing the probabilities that the geodesic will enter
each vacuum $i$ multiplied by the physical density of observations of
type $\mu$ in vacuum $i$, so this more general formula reduces to our
earlier result.

We thus reformulate the no-collapse scale factor measure as a local
measure.  By this we mean a measure that involves averaging over the
statistical ensemble defined by the different possible histories along
a single worldline, without necessarily assembling them into a global
geometry.  It would be natural to use the local formulation as the
general definition of the scale factor measure, since it can be
applied to collapsed regions without requiring additional rules.  

The causal-diamond measure is another example of a local measure in
this sense.  It differs from the no-collapse scale factor measure only
in two ways: (1) the transverse volume included along the geodesic is
not constant, but is set by the size of a causally connected region;
and (2) initial conditions are not determined by the causal diamond
measure, but are considered a logically independent question.  This
explains the pattern of similarities between the measures that has
emerged, but it also clarifies how they differ.

In Sec.~\ref{sec-inh}, we go beyond the no-collapse approximation and
investigate how the De Simone {\em et al.} prescription for collapsed
regions affects the formulae for probabilities in the scale factor
measure.  We find that it can be incorporated by a simple
substitution: Instead of using the physical density of observations,
what matters is the (potentially far greater) density those
observations {\em would have had\/}, if they had occured at the time
when the first structures formed that would later merge into the
objects hosting the observations.  In other words, we can incorporate
gravitational collapse by mapping each observation back to the
earliest time when a geodesic on which it lies began to collapse.  Our
own observations, for example, are thus condensed by inverse of the
expansion factor of our universe since the formation of the first dark
matter halos.

Since neither the inflationary era nor the early post-inflationary
universe contain collapsed regions, this modification has no bearing
on the results (1) and (2) above, concerning the youngness paradox and
the $Q$-catastrophe.  In Sec.~\ref{sec-cc}, we investigate how the
inclusion of gravitational collapse affects the prediction for the
cosmological constant.

We begin by reviewing the predictions for the cosmological constant
obtained in various measures: the observers-per-baryon prescription
(Sec.~\ref{sec-opb}), the causal diamond measure (Sec.~\ref{sec-cd})
and the no-collapse scale factor measure (Sec.~\ref{sec-noc}).  In
Sec.~\ref{sec-col}, we consider the scale factor measure, with the
prescription of Ref.~\cite{DGSV08} for collapsed regions.  We find that the
cosmological constant is not set by the timescale when observations
are made---as it is in the causal diamond measure and, apparently, in
Nature.  Rather, its value is controlled by the time scale of
structure formation.  This means that the scale factor measure
predicts a value that is up to 5000 times larger than the observed
value.  By adding more specific anthropic assumptions, the discrepancy
can be mitigated, but the observed value remains somewhat atypical.

If the primordial density contrast, $Q$, is allowed to vary, we find
that the preferred value of $\Lambda$ scales like $Q^3$, and the
associated probability like $Q^6$.  This means, for example, that a
universe with $Q$ three times as large, and $\Lambda$ 27 times as
large, is 729 times as likely.  It is difficult to see how anthropic
constraints or prior distributions for $Q$ can overcome a pressure so
strong.  In Sec.~\ref{sec-improv}, we discuss possible modifications
of the treatment of collapsed regions in the scale factor measure,
which might improve these problematic predictions.

In Sec.~\ref{sec-BB}, we investigate the probability for Boltzmann
brains in the scale factor measure.\footnote{A.~Vilenkin and collaborators have independently
  analyzed this question. We undertand that their results will appear
  simultaneously or in the near future.}.  Boltzmann brains are observers
that arise from rare thermal fluctuations, at a superexponentially
small rate per unit four-volume.  Some measures overcompensate for
this suppression by including superexponentially large regions of
empty de~Sitter space
\cite{Efs95,Vil95,Wei96,MarSha97,Vil04,GarSch05,Wei05} for every
region containing ordinary observers.  Then the vast majority of
observations are made by Boltzmann brains.  Since almost none of the
observations made by Boltzmann brains agree with our observations,
these measures are ruled out.  The causal diamond measure was shown in
Ref.~\cite{BouFre06b} to favor ordinary observers, if the decay rate
of every vacuum in the landscape is greater than the rate for
Boltzmann brains.  Recent evidence suggests that this nontrivial
condition may be satisfied in the string theory
landscape~\cite{FreLip08}.

Our local formulation of the scale factor measure makes it clear that
similar conditions will play a role in the scale factor measure.  The
transverse volume along the defining geodesic is the same in all
vacua, whereas in the causal diamond measure, it is set by the
cosmological constant.  Since the latter varies at most over an
exponentially, but not a superexponentially large range, this
difference is negligible in the context of Boltzmann brains.

The only remaining potential difference arises from the initial
conditions on the geodesic.  In the causal diamond measure, it was
reasonable to assume that these conditions do not pick out vacua with
unnaturally small cosmological constant (a necessary condition even
for Boltzmann brains), so the initial vacuum received no attention in
the analysis of Ref.~\cite{BouFre06b}.  In the scale factor measure,
however, the initial vacuum is the longest-lived de~Sitter vacuum,
which might well have a small cosmological constant.

Here, we refine the analysis of Ref.~\cite{BouFre06b} in two respects.
In Sec.~\ref{sec-ratio}, we include the contributions from the initial
vacuum to the number of Boltzmann brains.  In
Sec.~\ref{sec-conditions}, we demonstrate (under plausible assumptions
on the structure of the landscape) that the production rates of
different vacua will not invalidate our earlier criterion for the
dominance of ordinary observers, and we find that it is augmented only
by the condition that the initial vacuum be completely unable to
produce Boltzmann brains (independently of its decay rate).  We argue
that this condition is likely to be satisfied.  Thus, the scale factor
measure and the causal diamond measure are virtually equivalent for
the purpose of Boltzmann brains.

\section{Definition and rate equations}
\label{sec-defrate}

In this section we define the scale factor cutoff and review some of
its basic properties.

\subsection{Definition of the scale factor cutoff}
\label{sec-def}

One approach to regulating the infinities in eternal inflation is
through a smooth congruence of timelike geodesics orthogonal to a 
(nearly arbitrary) finite spacelike surface $\Sigma_0$.  The idea is 
to use the geodesic congruence to define a sequence of cutoff 
hypersurfaces $\Sigma_\eta$.  Then one computes the number $N_i$ of 
observations of type $i$, $O_i$, in the four-volume between $\Sigma_0$ 
and $\Sigma_\eta$.  The relative probability of two observations is 
defined by their relative abundance in the limit where the cutoff is 
taken away:
\begin{equation}
  \frac{{\cal P}(O_i)}{{\cal P}(O_j)}\equiv
  \lim_{\eta\to\infty}\frac{N_i}{N_j}
\label{eq-prob}
\end{equation}
A simple choice for constructing $\Sigma_\eta$ would be to follow each
geodesic for the same proper time. But the resulting measure is ruled
out at overwhelming confidence level, and independently of the details
of the underlying theory: It predicts that we should observe a much
hotter
universe~\cite{LinLin96,Gut00a,Gut00b,Gut04,Teg05,Lin07,Gut07,BouFre07}.

\subsubsection{The no-collapse scale factor measure}

A different cutoff on the congruence was recently defined by De Simone
{\em et al.}~\cite{DGSV08}, building on earlier work.\footnote{Scale
  factor time is considered in some
  of the earliest literature on (slow-roll) eternal inflation, which
  focusses on the distribution of different field
  values~\cite{LinLin94,GarLin94,GarLin94a,GarLin95}.  The measure
  defined by De Simone {\em et al.} is similar to the
  ``pseudo-comoving volume-weighted measure''~\cite{Lin06}, which
  stops short of an explicit definition of the probability of
  different observations, such as Eq.~(\ref{eq-prob}), Eq.~1 of
  Ref.~\cite{BouFre07} or Eq.~1 of Ref.~\cite{DGSV08}, and of a
  prescription for collapsing geodesics.  It shares some properties
  with other measures~\cite{EasLim05,GarSch05,VanVil06,Van06,Vil06},
  in which the scale factor (in the guise of approximately equivalent
  formulations) regulates bubble abundances but different (or no)
  cutoffs regulate the number of observers per bubble.}  In this
proposal $\Sigma_\eta$ is, roughly, a surface of constant local scale
factor.  The {\em scale factor time} is defined by integrating the
local expansion rate relative to infinitesimally nearby geodesics,
along each geodesic:
\begin{equation}
  \eta(\mathbf{x},t) \equiv \int_0^t \frac{\theta}{3} dt'~, 
\label{eq-def}
\end{equation}
where $t$ is the proper time along the geodesic labeled by
$\mathbf{x}$, $\theta=\nabla^\mu\zeta_\mu$ is the expansion, and
$\zeta_\mu$ is the 4-velocity vector field tangent to the congruence.
The {\em local scale factor} is defined by
\begin{equation}
  A(\mathbf{x} ,t)\equiv \exp \eta(\mathbf{x} ,t)~.
\label{eq-capa}
\end{equation}
Intuitively, these quantities measure the growth of a local volume
element $\delta V$ spanned by infinitesimally nearby
geodesics~\cite{Wald}:
\begin{equation}
  \theta = 3\frac{d\eta}{dt}=3\frac{dA/dt}{A}=\frac{d(\delta
    V)/dt}{\delta V}~.
\label{eq-volume}
\end{equation}

\subsubsection{Collapsed regions and other ambiguities}
\label{sec-ambiguities}

However, $\Sigma_\eta$ cannot be defined simply as a surface of
constant $\eta$ or $A$.  In collapsing regions, such as pockets with a
negative cosmological constant, or structure forming regions,
geodesics will cease to expand and begin to approach each other.  Then
$\theta<0$, and by Eq.~(\ref{eq-def}), the scale factor time decreases
locally towards the future.  Unless a singularity is encountered, the
focusing theorem guarantees that geodesics will eventually encounter
caustics: they will intersect with infinitesimally neighboring
geodesics.  Thereafter, they begin to expand again, and the scale
factor time increases once more.

This introduces two ambiguities: First, a given scale factor need not
be reached precisely once along each geodesic.  It may never be
reached, or it may be reached more than once.  Which (if any)
occurrence defines the actual cutoff?  Second, a given event may be
threaded by more than one geodesic.  Should we count such events once,
or multiple times?

To resolve these ambiguities, De Simone {\em et al.} \cite{DGSV08}
propose that an event should be included ``if it lies on any geodesic
prior to the first occurrence'' of the specified cutoff on that
geodesic.  In other words, $\Sigma_\eta$ is the hypersurface that
maximizes the four-volume $V_4$ of the congruence subject to the
constraint that at every point $p\in V_4$ lies on at least one
geodesic at scale factor time less than $\eta$.

\begin{figure}[t!]
\begin{center}
\includegraphics[scale=.32]{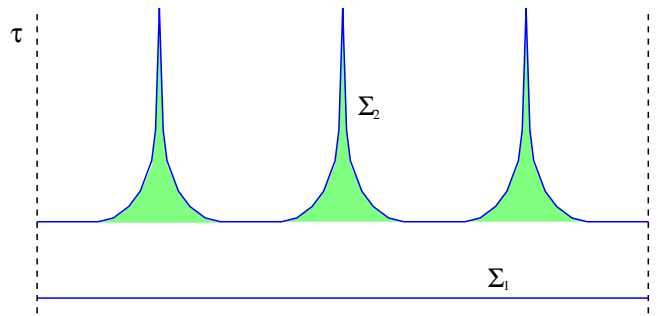}
\caption{Evolution of the scale factor cutoff during structure
  formation. The constant scale factor time surface $\Sigma_1$ lies in
  the early, approximately homogeneous universe and coincides with a
  surface of constant FRW time $\tau$. As density perturbations grow,
  some geodesics decouple from the Hubble flow, stop expanding, and
  become trapped in collapsed regions such as galaxies. If the scale
  factor cutoff exceeds the largest scale factor ever reached along
  such geodesics, then the rule of De Simone {\em et al.} requires
  that their entire future evolution be included. (A similar result
  obtains if the local formulation of the scale factor measure is used
  as a general definition; see Sec.~5.4.) Therefore, the later cutoff
  surface $\Sigma_2$ no longer agrees everywhere with a constant FRW
  time surface; it includes the entire future of collapsed regions
  (green/gray), which show up as spikes since
  the figure is drawn in comoving coordinates.}
\label{fig-spikes}
\end{center}
\end{figure}

This definition implies that if the cutoff is never reached along some
geodesic, all events on the geodesic are included (no future cutoff).
The formulation of De Simone {\em et al.} suggests, moreover, that any
event should be counted at most once; we will adopt this definition
here. Note that with this definition, $\Sigma_\eta$ need not be a
spacelike hypersurface. Rather, it becomes timelike near collapsing
geodesics, spiking up towards the future, as shown in
Fig.~\ref{fig-spikes}.

The prescription for collapsed regions chosen by De Simone {\em et
  al.} is not the only possible one. Indeed, we will find that it has
phenomenological disadvantages (Sec.~\ref{sec-cc}), and we will
suggest other possible definitions. In particular, the local
formulation of the scale factor measure (Sec.~\ref{sec-local}) applies
without modification both in collapsing and expanding regions.
However, we will not explore these alternatives in detail; we focus on
the definition of Ref.~\cite{DGSV08} in this paper.

There is another ambiguity, however, which has yet to be resolved.
When a new vacuum with lower cosmological constant is formed by the
Coleman-DeLuccia process, there is a quantum region, where the
geometry (in the instanton approximation) jumps sharply and the
continuation of geodesics is not well-defined.  For downward
tunnelings, one can reasonably ignore this problem, since the initial
radius of the bubble wall can be much smaller than the de~Sitter
radius, so only a negligible fraction of geodesics entering the new
bubble pass through the bubble interior at the time of nucleation.
Moreover, geodesics quickly become comoving in regions with slow-roll
inflation, which are of the greatest interest.  

De Sitter vacua may also tunnel upward, to vacua with larger
cosmological constant. Even though upward tunnelings are very
suppressed, the scale factor measure implies that majority of
observers live on world lines which have been through upward
tunnelings~\cite{Sch06,Sch08}. But there is no approximate classical
geometry describing an upward tunneling. There is no natural way,
therefore, of determining the tangent vectors of the portion of the
geodesic congruence entering the new bubble.

For the purpose of deriving the rate equations, in
Sec.~\ref{sec-rate}, we will follow Ref.~\cite{GarSch05} and sidestep
this issue by making mathematically convenient assumptions. Once we
count observers in Sec.~\ref{sec-OO}, the ambiguity will not re-enter:
we are only interested in the most recent tunneling, which is
necessarily downward for producing ordinary observers.

\subsection{Rate equations and solutions}
\label{sec-rate}

Before we can count observers in vacuum $i$, we will need to know the
number $n_i(\eta)$ of such bubbles below the cutoff $\eta$.  This, in
turn, requires knowing the physical volume $V_i(\eta)$ occupied by
every vacuum.  The evolution and distribution of these volumes is
determined by the rate matrix $\Gamma_{ij}$ describing the number of
bubbles of vacuum $i$ forming per unit four volume of vacuum $j$.

For nonterminal vacua (i.e., those with positive cosmological constant
$\Lambda_i$), the following definitions will be convenient: the
expansion at late times,
\begin{equation}
H_i\equiv \left(\frac{\Lambda_i}{3}\right)^{1/2}~;
\end{equation}
the total decay rate of vacuum $i$ per unit four-volume,
\begin{equation}
\Gamma_j\equiv\sum_i\Gamma_{ij}~;
\end{equation}
the dimensionless decay rate from vacuum $j$ to $i$,
\begin{equation}
\kappa_{ij}\equiv\frac{4\pi}{3}\frac{\Gamma_{ij}}{H_j^4}~;
\label{eq-dimensionless}
\end{equation}
the total dimensionless decay rate of vacuum $j$,
\begin{equation}
  \kappa_j \equiv \sum_i\kappa_{ij}~,
\end{equation}
and the branching ratio matrix
\begin{equation}
\beta_{ij} \equiv \kappa_{ij}/\kappa_j~.
\label{eq-branch}
\end{equation}

During a time interval $d\eta$, the volume $V_i$ in vacuum $i$ will
increase due to intrinsic expansion, due to the production of new
bubbles of type $i$, and due to the expansion of the bubble walls into
the parent vacua.  It will decrease due to decay into other vacua, and
due to the growth of such bubble walls after they are produced.

Treating the motion of domain walls in detail is cumbersome, and it is
unnecessary if all metastable vacua are long-lived ($\kappa_i\ll 1$).
This is a reasonable assumption, since $\kappa_i\gg 1$ conflicts with
the notion of a vacuum, and $\kappa_i\sim 1$ requires fine tuning.
Such vacua will be too rare in the landscape to play an important
dynamical role.

Thus, most of the four-volume of each bubble type is empty de~Sitter
space, and we can neglect transient effects right after bubble
nucleation.  One transient is the period between bubble creation and
vacuum domination.  Therefore, the expansion $\theta$ of geodesics in
vacuum $j$ can be approximated by the Hubble constant at late times,
$\theta\approx H_j\equiv (\Lambda_i/3)^{1/2}$.  The four-volume in
vacuum $j$ added during the scale factor time $d\eta$ is thus $V_j
H_j^{-1} d\eta$.

Another transient is the bubble wall expansion.  A bubble nucleated at
the time $\eta^{\rm nuc}$ will eventually occupy a comoving volume in
the geodesic congruence that would (in the absence of the decay) have
originated from a ball of physical radius $H_j^{-1}$ in the parent
vacuum $j$ at the time $\eta^{\rm nuc}$.  This asymptotic comoving
size is reached, to accuracy of order $\exp(\eta-\eta^{\rm nuc})$,
after only a few units of scale factor time.  Thus we make a small
error by anticipating this growth: We ascribe a physical volume
$4\pi/3H_j^3$ to the new vacuum $i$ already at the nucleation time
$\eta^{\rm nuc}$, and in exchange neglect the bubble wall
growth~\cite{GarSch05}.\footnote{This behavior of the bubble wall
  applies only to downward transitions.  During upward transitions,
  the behavior of the congruence defining the scale factor measure is
  not well-defined in semi-classical gravity.  Following
  Ref.~\cite{GarSch05}, we will choose {\em ad
    hoc\/} to use the same rule in this case.}

With these approximations, vacuum $i$ gains
\begin{equation}
  d^+V_i= 3 V_i\, d\eta +  \sum_j\left(V_j\frac{d\eta}{H_j}\right)
    \Gamma_{ij}\left(\frac{4\pi}{3H_j^3}\right)~.
\end{equation}
and loses
\begin{equation}
d^-V_i=\sum_j\left(V_i\frac{d\eta}{H_i}\right)
\Gamma_{ji}\left(\frac{4\pi }{3H_i^3}\right)~.
\end{equation}
in physical volume per scale factor time.  Combining inflow and
outflow yields the Fokker-Planck equation
\begin{equation}
  \frac{dV_i}{d\eta}= 3 V_i-\kappa_iV_i+\sum_j\kappa_{ij}V_j~.
\end{equation}

Ref.~\cite{GarSch05} rigorously derives the solution to this equation.
Only the behavior of de~Sitter vacua will be relevant here.  For
generic initial conditions with some support in de~Sitter vacua, the
solution approaches attractor behavior at late times.  The volume in
de~Sitter vacuum $i$ is
\begin{equation}
V_i(\eta) = C\, s_i\, e^{\gamma\eta}~.
\label{eq-volumes}
\end{equation}
Here, $C$ is a constant with the dimension of volume that depends on
the initial conditions but drops out in all normalized probabilities;
\begin{equation}
\gamma\equiv 3-q~;
\end{equation}
$q$ is the smallest-magnitude negative eigenvalue of the dimensionless
flow matrix $\kappa_{ij}-\delta_{ij}\kappa_i$; and $s_j$ is the
associated eigenvector:
\begin{equation}
  \kappa_{ij} s_j = (\kappa_i-q) s_i\equiv p_i~,
\label{eq-eigen}
\end{equation}
where we have defined the vector $p_i$ for later convenience.

To exponentially good approximation~\cite{SchVil06}, the eigenvector
is dominated by the longest-lived de~Sitter vacuum, $*$:
\begin{equation}
s_j\approx\delta_{j*}~;
\label{eq-sstar}
\end{equation}
and its eigenvalue is equal to the dimensionless decay rate of this
vacuum,
\begin{equation}
q\approx \kappa_*\ll 1
\label{eq-qkappa}
\end{equation}
Thus, $q$ is exponentially small in a realistic landscape.---By
Eq.~(\ref{eq-prob}), this attractor solution is all we need to compute
probabilities.

A number of other results will be useful below, and we collect them
here.  The expected number of times a worldline starting in vacuum
$o$ vacuum will pass through vacuum $i$ can be obtained by
summing over the whole branching tree~\cite{Bou06}:
\begin{equation} 
  e_{io} = \delta_{io} +
  \sum_{{\rm paths~from}~o~{\rm to}~i} \beta_{i i_n}~
  \beta_{i_n i_{n-1}}~ ... ~\beta_{i_1 o}~,
\label{eq-panth}
\end{equation}
where $i_1...i_n$ are intermediate vacua connecting $o$ and $i$, and
the branching ratios $\beta_{ij}$ were defined in
Eq.~(\ref{eq-branch}).  This can be written in matrix form as
\begin{equation}
  e_{io} = \bigg(\sum_{n=0}^{\infty}\bm{\beta}^n\bigg)_{io}~,
\label{eq-ei}
\end{equation}
In the scale factor measure, the initial vacuum is the $*$ vacuum, and
we will denote
\begin{equation}
  e_i\equiv e_{i*}
\end{equation}
If the initial vacuum has relatively small cosmological constant, as
one might expect for the longest-lived vacuum, then it is unlikely to
be re-entered later in the decay chain.  Then the sum in
Eq.~(\ref{eq-ei}) converges rapidly; in particular,
\begin{equation}
  e_*\approx 1~.
\label{eq-estar}
\end{equation}

In Ref.~\cite{BouYan07}, the vectors $p_i$ and $e_i$ were shown to be
closely related:
\begin{eqnarray}
  p_i &=&     q \bigg(\sum_{n=1}^{\infty}\beta^n\bigg)_{ij}s_j 
  \nonumber \\
  & \approx & q \bigg(\sum_{n=1}^{\infty}\beta^n\bigg)_{i*}~.
  \nonumber \\
  &\approx & q e_i~~~~~(i\neq *)
\label{eq-piei}
\end{eqnarray}
Note, however, that $p_*\neq qe_*$.

\section{The cutoff hypersurface in a homogeneous bubble}
\label{sec-one}

In this subsection, we investigate the evolution of the cutoff
hypersurface in a single bubble of an arbitrary vacuum, nucleated
inside a parent vacuum with positive cosmological constant $3H^2$.  We
will drop the index $i$ while discussing a single vacuum.  For now, we
will ignore inhomogeneities, such as density perturbations and local
gravitational collapse.

Assuming that the tunneling process is very rare, the parent vacuum
will have existed for a long time before the nucleation event.  This
implies~\cite{GarGut06} that we can take the geodesic congruence to be
comoving in the flat slicing of de Sitter space,
\begin{equation}
ds^2=-dt^2+e^{2Ht}(dr^2+r^2d\Omega^2)~.
\end{equation}
We can think of $r$ as labelling a particular shell in the congruence.
The geodesics between $r$ and $r+\delta r$ span a volume element
\begin{equation}
  \delta V(t,r)=\left.\frac{dV}{dr}\right|_{t=\rm{const}}
  \delta r= 4\pi e^{2Ht} r^2 \delta r
\label{eq-vtr}
\end{equation}
in the parent vacuum.

In the homogeneous approximation, the metric inside the bubble is
described by an open FRW geometry
\begin{equation}
\label{eq-FRW}
ds^2 = - d\tau^2 + a^2(\tau) (d \xi^2 + \sinh^2\xi\, d \Omega_2^2)~.
\end{equation}
The comoving geodesics from the host vacuum continue into the bubble
along nontrivial trajectories, defining a second coordinate system
$(t,r)$ inside the bubble, where $t$ is the proper time along the
geodesic passing through an event, and $r$ is the radial position the
geodesic had before entering the bubble.  We will now review the
coordinate transformation between $(\tau,\xi)$ and $(t,r)$, derived in
Ref.~\cite{BouFre07}.

Without loss of generality, one can choose coordinates so that the
bubble is nucleated at $t=0,r=0$.  Consider a comoving geodesic which
passes through the event $t = 0, r$.  If after a proper time $t$ (as
measured along the geodesic) it has passed into the bubble, its
coordinates $(\tau,\xi)$ will be
\begin{eqnarray}
  \xi(t,r)  &=& -\log(1- Hr) +  {\cal O}(\frac{1}{H\tau}) 
  \nonumber \\
  \tau(t,r) &=& t -\frac{\xi+e^{-\xi}-1}{H}+{\cal O}(\frac{1}{H\tau})~.
  \label{eq-trans}
\end{eqnarray}
Note that $H$ in these formulas refers to the constant expansion rate
of the parent de~Sitter space.  The above equations are valid for
geodesics that have spent more than a proper time $H^{-1}$ inside the
bubble.\footnote{We also assume that the cosmological 
constant inside is much smaller than outside.  Corrections due to such 
approximation are not included in Eq.~(\ref{eq-trans}).  However, when 
the inside and outside cosmological constants are the same, one can 
derive an exact formula which is equivalent to Eq.~(\ref{eq-time}) for
all physical questions we considered in this paper.} Physically, this 
result shows that the geodesics become comoving in the open 
coordinates after one outside Hubble time, and thereafter the proper 
time along the geodesics increases at the same rate as the open FRW 
time.

To construct the cutoff hypersurface, one must determine the scale
factor time $\eta$ as a function of the bubble coordinates $\tau$ and
$\xi$.   By Eq.~(\ref{eq-volumes}),
\begin{equation}
  \eta(\eta^{\rm nuc},\tau,\xi) = \eta^{\rm nuc}+ 
  \frac{1}{3}\log\left(\frac{\delta 
      V[r(\tau,\xi),t(\tau,\xi)]}{\delta V(r,0)}\right)~,
\label{eq-volrat}
\end{equation}
where $\eta^{\rm nuc}$ is the scale factor time at $t=0$, when the bubble is
nucleated.  The volume element $\delta V$ orthogonal to the geodesics
lies on a hypersurface of constant proper time $t$~\cite{Wald}.  By
Eq.~(\ref{eq-FRW}),

\begin{eqnarray} 
  && \delta V = \nonumber \\
  && 4\pi a^2(\tau) \sinh^2 \xi 
  \left[a^2(\tau)\left(\frac{d \xi}{dr}\right)_{t={\rm const}}^2 
                -\left(\frac{d\tau}{dr}\right)_{t={\rm const}}^2
     \right]^{1/2}  \delta r \nonumber \\
  && = 4\pi a(\tau)^2 \sinh^2 \xi\, 
  \left[a(\tau)^2-\left(\frac{1-e^{-\xi}}{H}\right)^2\right]^{1/2} 
  H e^\xi\, \delta r~, 
\end{eqnarray}

where we have used Eq.~(\ref{eq-trans}).  We are interested in events
much later than the outside Hubble time, so $a^2 \gg H^{-2}$, and we
can drop the second term in the square root:
\begin{equation}
  \delta V = 4\pi H a^3(\tau) e^{\xi} \sinh^2 \xi\, \delta r
\end{equation}
After using Eq.~(\ref{eq-trans}) to eliminate $r$ from
Eq.~(\ref{eq-vtr}), Eq.~(\ref{eq-volrat}) yields the scale factor time
inside the bubble:
\begin{equation}
  \eta(\eta^{\rm nuc},\tau,\xi)= \eta^{\rm nuc} +
  \log\left[H a(\tau)\right]+
  \frac{2}{3}\left[\xi+\log\left(\cosh\frac{\xi}{2}\right)\right]~.
\label{eq-time}
\end{equation}

\section{Counting ordinary observers}
\label{sec-OO}

In this section, we apply the scale factor cutoff to counting
observations in an eternally inflating multiverse with multiple vacua.
We exclude, for now, observations resulting from violations of the
second law (Boltzmann brains, treated in Sec.~\ref{sec-BB}).
Initially, we will imagine that all observations (if any) in a bubble
of type $i$ are made instantaneously\footnote{In our investigation of
  the proper time cutoff~\cite{BouFre07}, information about the
  temporal distribution of observations, $f_i(\tau)$, was crucial to
  demonstrating the youngness paradox.  In the scale factor measure,
  however, there is no youngness paradox~\cite{DGSV08}, and for the
  purposes of this paper, it suffices to use only
  $g_i\equiv\int_0^\infty f_i(\tau)d\tau$ and $\tau^{\rm obs}_i\equiv
  g_i^{-1}\int_0^\infty f_i(\tau) \tau d\tau$ as input parameters.} at
the FRW time $\tau^{\rm obs}_i$ after the vacuum is produced, with
number density $\rho_i^{\rm obs}$ per unit physical volume:
\begin{equation}
  dN_i = \rho_i^{\rm obs} dv_{\rm ph}~.
\label{eq-nvol}
\end{equation}
We will assume, moreover, that these parameters do not depend on the
parent vacuum from which $i$ is entered.  These assumptions will
simplify our treatment, and it will be easy to drop them in the end
and state a more general result, Eq.~(\ref{eq-fatline}).

Our treatment will go beyond that of De Simone {\em et al.}
~\cite{DGSV08} in that we do not approximate the bubble interior as
flat and homogeneous.  Our detailed treatment of collapsing regions,
in Sec.~\ref{sec-inh}, has significant implications, invalidating some
of the conclusions of Ref.~\cite{DGSV08} (see Sec.~\ref{sec-cc}).

\subsection{Counting observations in the no-collapse approximation}
\label{sec-homcount}

Now we will compute the total number of observations, $N_i(\eta)$,
performed in vacua of type $i$ below the cutoff, $\eta$.  By
Eq.~(\ref{eq-nvol}), this amounts to computing the total physical
volume of the $\tau= \tau_i^{\rm obs}$ hypersurfaces below the cutoff.
In any single bubble, a shell of comoving radius $\xi$ and width
$d\xi$ contributes a physical volume
\begin{equation}
  dv^{\rm ph}_i=4\pi (a_i^{\rm obs})^3 \sinh^2\xi\, d\xi~,
\end{equation}
but only if the bubble was nucleated early enough
for this volume to be included below the cutoff.  

By Eq.~(\ref{eq-time}), the latest nucleation time that allows
observers at radius $\xi$ to contribute is
\begin{equation}
  \eta_{ij}^{\rm nuc}(\eta,\xi) = \eta -
  \log\left[H_j a_i^{\rm obs}\right]- 
  \frac{2}{3}\left[\xi+\log\left(\cosh\frac{\xi}{2}\right)\right]~.
\end{equation}
Note that this time depends on the parent vacuum, $j$.  Hence,
\begin{equation}
  N_i(\eta)=\sum_J\int_0^\infty d\xi\, 
  n_{ij}\!\left[\eta_{ij}^{\rm nuc}(\eta,\xi)\right]\, \frac{dv_{\rm
      ph}}{d\xi}(\xi)\, \rho_i^{\rm obs},
\end{equation}
where $n_{ij}(\eta')$ is the total number of bubbles of type $i$
nucleated inside vacuum $j$ by the time $\eta'$.

Using
\begin{equation}
dn_{ij}=\Gamma_{ij} V_j d\eta/H_j
\end{equation}
Eqs.~(\ref{eq-dimensionless}) and (\ref{eq-volumes}) imply
\begin{equation}
  n_{ij}(\eta_{ij}^{\rm nuc})= \frac{3C}{4\pi\gamma} H_j^3 \kappa_{ij} s_j 
  \exp(\gamma\eta_{ij}^{\rm nuc})~.
\label{eq-nsimple}
\end{equation}

Combining the above equations, we find
\begin{eqnarray} 
  && N(\eta) = \nonumber \\
  &&\frac{3C}{4\pi\gamma} \left[\int_0^\infty d\xi\,  4\pi\, 
    \sinh^2\xi\, (\exp\xi)^{-2\gamma/3} 
    \left(\cosh\frac{\xi}{2}\right)^{-2\gamma/3}\right]e^{\gamma\eta} 
  \nonumber\\
  && \times (a_i^{\rm obs})^q \left(\sum_j H_j^q \kappa_{ij} s_j\right)
  \rho_i^{\rm obs} ~.
\end{eqnarray} 
The $\xi$ integral is clearly convergent, with support concentrated
near $\xi\sim O(1)$.  (For small $q$, its value approaches $4\pi/3$.)
Physically, this shows that mainly the central curvature volume of the
infinite open bubble geometry contributes in the scale factor measure.
Therefore, the interesting results of Garriga, Guth, and Vilenkin
\cite{GarGut06} regarding worldlines at large $\xi$ are not relevant
in this measure.

Since $q$ is exponentially small, $H_j^q$ and $(a_i^{\rm obs})^q$ can
safely be neglected.  By the discussion at the end of
Sec.~\ref{sec-defrate}, $\kappa_{ij} s_j=q e_i$.  Thus we obtain the
probability
\begin{equation}
  {\cal P}_i\propto e_i\, \rho_i^{\rm obs} ~.
\label{eq-main}
\end{equation}
for observing vacuum $i$.  The ``$\propto$'' notation indicates that
the universal factor $\approx Cq e^{\gamma\eta}/\gamma$ has been
dropped (since, by Eq.~(\ref{eq-prob}), it does not affect relative
probabilities), though the probabilities have not been normalized.

Eq.~(\ref{eq-main}) immediately implies several key properties of the
scale factor measure:
\begin{itemize}

\item The probability for observing vacuum $i$ is simply the product
  of the number of times a typical worldline can be expected to enter
  $i$-bubbles, $e_i$, and the density, $\rho_i^{\rm obs}$, of
  observations per physical volume on the homogeneous timeslice
  $\tau_i^{\rm obs}$ at which they are performed.

\item The probability of an observation depends on its FRW time,
  $\tau^{\rm obs}$, and on the FRW expansion factor at that time,
  $a^{\rm obs}$, {\em only\/} through $\rho^{\rm obs}$.

\item As a special case, we recover an important result of De Simone
  {\em et al.}~\cite{DGSV08}: Like in the causal diamond measure, the
  probability of a vacuum is insensitive to the volume expansion
  factor during slow-roll inflation.  This is a desirable property,
  because it avoids the
  ``Q-catastrophe''~\cite{GarVil05,FelHal05}---the overwhelming
  pressure towards extreme (and counterfactual) inflationary parameter
  values that results when exponential volume factors are rewarded.
  Moreover, this property makes it conceivable that inflation was
  short enough to allow subtle signatures of the preceding era to
  survive, such as detectable curvature~\cite{FreKle05}.

\item Eq.~(\ref{eq-main}) also captures another important result of
  Ref.~\cite{DGSV08}, which we will examine in Sec.~\ref{sec-cc}: the
  probability of vacua where vacuum energy comes to dominate before
  $\tau_i^{\rm obs}$ is exponentially suppressed, because the matter
  density, $\rho_{{\rm matter},i}^{\rm obs}$, will have been diluted
  by the accelerated expansion.  This, too, appears to replicate a
  success of the causal diamond measure: the suppression of moderately
  large values of the cosmological constant $\Lambda$, which are
  larger than the observed value but too small to affect the number of
  observers per matter mass.  We will find, however, that this
  apparent success is an artifact of the no-collapse approximation.

\end{itemize} 

Eq.~(\ref{eq-main}) trivially generalizes to the case where
observations are made at different different times $\tau_\mu$,
$\mu=1,2,3\ldots$ The total probability to observe vacuum $i$ is
\begin{equation}
{\cal P}_i\propto e_i \sum_\mu \rho^{\rm obs}_i(\tau_\mu)~,
\label{eq-sum}
\end{equation}
where $\rho^{\rm obs}_i(\tau_\mu)$ is the density of observations
taking place at $\tau_\mu$.  In the continuous case,
\begin{equation}
  {\cal P}_i\propto e_i \int_0^\infty d\tau\,
  \frac{d\rho^{\rm obs}_i}{d\tau}~,
\label{eq-cont}
\end{equation}
where the integrand is the number of observations made in the FRW time
interval $(\tau, \tau+d\tau)$, per unit physical volume of the
hypersurface $\tau$.  If more differentiated observations are
performed (for example, local conditions like temperature, rather than
just a determination of the vacuum), $\rho_i$ can be given additional
indices or arguments.

\subsection{The no-collapse scale factor measure as a local measure}

\label{sec-local}

In the no-collapse approximation, the scale factor measure can be
reformulated as a local measure.  By this we mean that the measure can
be defined by averaging over an ensemble of individual worldlines,
instead of constructing a particular global spacetime.  (The causal
diamond measure is an example of a local measure.)

\begin{figure}[t!]
\begin{center}
\includegraphics[scale=.34]{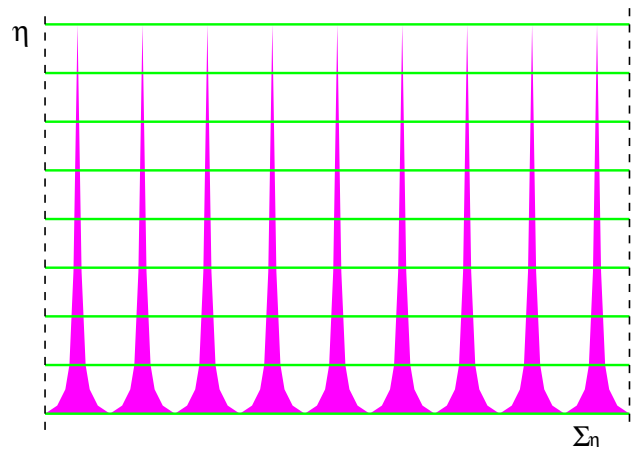}
\caption{The green (light shaded) slices are surfaces of constant 
scale factor time, $\Sigma_\eta$; they have fixed comoving size but 
increasing physical volume.  Fat geodesics (purple, dark shaded) have 
fixed physical width and thus decreasing comoving size.  If the 
initial slice is chosen in the attractor regime, the fat-geodesics 
define a representative finite sample of the total four-volume.  Thus, 
the results of the scale factor measure can be reproduced by following 
a single geodesic of fixed width, starting in the longest-lived 
metastable vacuum.}
\label{fig-local}
\end{center}
\end{figure}

This can be seen by the following argument, illustrated in
Fig.~\ref{fig-local}.  The predictions of the scale factor measure are
dominated by the late-time attractor behavior of the universe.  (An
infinite number of observations are produced during this era, while
only a finite number are produced earlier.)  Thus, we may as well pick
a late-time surface, $\Sigma_\eta$ with $\eta$ very large, and choose
it as our initial surface, $\Sigma_0$.  In other words, we redefine
$\eta\to\eta-\Delta\eta$, $\Delta\eta\approx\infty$, making sure we
already start in the asymptotic regime.  This cuts out irrelevant
transients and leads to a very simple picture.

The volume occupied by long-lived metastable vacua on the late-time
attractor hypersurface $\Sigma_0$ is dominated by empty de~Sitter
regions, with volume fraction $s_i$ allocated to vacuum
$i$~\cite{GarSch05}.  We can choose $\Sigma_0$ to be as large as we
like, guaranteeing that it can be allocated the correct attractor
volume fractions to arbitrary accuracy.

Now let us begin evolving forward along the congruence orthogonal to
$\Sigma_0$.  The scale factor measure instructs us to count
observations in the four-volume thus generated, taking ratios as
$\eta\to\infty$.  Let us instead consider a reduced four-volume,
$\tilde V_4$, which is finite, by dividing the initial surface
$\Sigma_0$ into volume elements $dV$.  As usual, we follow the
geodesic orthogonal to each volume element.  But we do not increase
$dV$ as the universe expands.  This creates a family of ``fat
geodesics''.  At the time $\eta$, the fat geodesics occupy only a
fraction $e^{-3\eta}$ of the total volume of $\Sigma_\eta$.  But for
every worldline, no matter in what type of region it started, the
missing volume is exactly the same, $1-e^{-3\eta}$.

Because all fat geodesics expand by the same volume factor, the volume
they do occupy on $\Sigma_\eta$ is a perfectly faithful sample of the
hypersurface.  (This is the crucial point---note, for example, that it
would not hold in the attractor regime of the proper time measure.)
Moreover, because this is true for every time interval $d\eta$, the
four-volume swept out by the fat geodesics out is statistically
equivalent to the four-volume between $\Sigma_0$ and $\Sigma_\eta$.

Therefore, reducing to $\tilde V_4$ will not affect probabilities:
\begin{equation}
\frac{{\cal P}(O_i)}{{\cal P}(O_j)}\equiv
\lim_{\eta\to\infty}\frac{N_i}{N_j}=
\frac{N_i(\tilde V_4)}{N_j(\tilde V_4)}~,
\end{equation}
where $V_4$ is obtained by following every geodesic to the asymptotic
future.\footnote{The no-collapse approximation does not allow us to
  consider negative cosmological constant regions, but for the
  purposes of this argument we can set $\Lambda_i\to 0$ for all vacua
  with $\Lambda_i<0$.}  Instead of thinking about the actual
collection of fat geodesics anchored on a large hypersurface, we may
equivalently consider a statistical ensemble of single geodesics, with
initial conditions weighted by the distribution of regions on a
late-time attractor surface.  Neglecting the rare regions occupied by
recently nucleated bubbles, this means starting out with de~Sitter
vacuum $i$ with probability $s_i$.  Because the vector $s_i$ is
dominated by the $*$ vacuum, to a good approximation~\cite{SchVil06},
this means starting the geodesic in the longest-lived metastable
vacuum, $*$.

Thus, we reproduce the scale factor measure by following a single
geodesic starting in $*$.  The worldline evolves according to local
dynamical laws, witnessing the decay of vacua and perhaps the
production of observers, until it ends up in a terminal vacuum. We
``fatten'' the worldline, giving it a {\em fixed physical
  cross-section\/}, a volume element $dV$ orthogonal to the worldline.
Finally, we compute the expectation value (ensemble average) of the
differential number of observations of type $\mu$, $dN_\mu$, in the
resulting four-volume:
\begin{equation}
{\cal P}_\mu=\langle \rho_\mu\rangle ~,
\label{eq-fatline}
\end{equation}
where, for a given worldline in the ensemble,
\begin{equation}
\rho_\mu = \frac{dN_\mu}{dV} = \int dt \frac{dN_\mu}{dV dt}
\label{eq-rhomu}
\end{equation}

We showed in Sec.~\ref{sec-one} that geodesics quickly become comoving
after entering a new bubble.  Thus, the volume element $dV$ lies
inside a constant-$\tau$ hypersurface of each FRW bubble universe.
With the approximations used in Sec.~\ref{sec-homcount}, therefore,
$\rho_\mu$ is identical to the physical density of observers,
$\rho_i^{\rm obs}$, in vacuum $i$.  The factor $e_i$ is captured by
averaging over different worldlines in the ensemble~\cite{BouYan07},
so Eq.~(\ref{eq-fatline}) reduces to Eq.~(\ref{eq-main}) as a special
case.  The more general Eqs.~(\ref{eq-sum}) and (\ref{eq-cont}), too,
are special cases of Eq.~(\ref{eq-fatline}).

It would be interesting to use Eq.~(\ref{eq-fatline}) as the {\em
  defining\/} equation of a measure.  This would have a number of
formal advantages.  One geodesic carries much less geometric
information than a whole congruence, so we face fewer ambiguities
about the treatment of upward jumps.  Moreover, Eq.~(\ref{eq-fatline})
can be applied without modification in collapsing regions, whereas the
formulation based on integrating the expansion along a geodesic
congruence requires an additional, {\em ad hoc\/}, prescription to
deal with such regions.  Finally, the use of Eq.~(\ref{eq-fatline}) to
define a measure liberates us from the initial conditions $s_i$ picked
out by the attractor regime of the geodesic congruence.  Starting the
fat geodesic with initial conditions that favor Planck-scale vacua,
for example, would avoid the potential ``staggering problem''
associated with the enormous suppression of the upward jumps from the
dominant vacuum $*$.

\subsection{Proper treatment of collapsed regions}
\label{sec-inh}

At least in our own bubble, some observations are made in collapsed
regions.  The FRW metric, Eq.~(\ref{eq-FRW}), does not capture the
local geometry of such regions, but provides only an average over
scales on which the universe can be considered homogeneous.  Hence,
this metric cannot be used to compute the local scale factor, $A$, and
the scale factor time, $\eta$, along geodesics entering collapsed
regions such as galaxies.  Despite the similar names, the FRW scale
factor $a$ and the local scale factor $A$ are two completely different
objects.

Geodesics that end up in dark matter haloes will, by definition, have
ceased to expand, and decoupled from the Hubble flow, before the halo
formed.  After reaching a maximum expansion, they will have turned
around and collapsed, with $A$ and $\eta$ decreasing during the
collapse.  These geodesics will reach observers, but their maximum
scale factor $A_{\rm max}$ will be unrelated to the FRW scale factor
at the time $t_{\rm obs}$, $a_{\rm obs}$.  Rather, it will be related
to the FRW scale factor at the time of structure formation, $a_{\rm
  NL}$.

Consider a simple, spherically symmetric model.  A dark matter halo
forms by the collapse of a spherical overdensity.  Then baryons fall
into its gravitational well, cool, and condense in the center.
Eventually the gas fragments into stars.

Let us focus on the dark matter particles that end up in the halo.
Initially, each particle follows one of the geodesics in the
congruence that defines the scale factor cutoff.  The maximum scale
factor is achieved at the time of turnaround, when all the dark matter
particles are momentarily stationary.  Note that it is smaller than,
but on the order of, the FRW scale factor $a$ at the time of
turnaround.

After the turnaround, the particles fall towards the center of the
halo.  Depending on interactions, the particles will eventually stop
following the geodesics, but in the spherical model, the geodesics
remain very simple.  They will focus in the center of the halo, and
begin expanding again, back out to the turnaround radius; this pattern
will be repeated indefinitely.  The entire congruence of geodesics
will keep oscillating about the center of the halo, with an amplitude
given by the turnaround radius.  This radius is twice the virial
radius, and perhaps ten times the eventual galaxy radius.  Therefore,
the congruence continues to thread the galaxy at all times.
Crucially, it will capture observers independently of how long it
takes to form them.\footnote{In the spherical idealization with an empty shell between overdense geodesics and the homogeneous background, there would be a small fraction of events that are missed by the congruence 
near the time when it is in focus. This would not be an important effect quantitatively, and will be absent in a realistic congruence.}

With a more realistic structure formation model, the situation would
appear to be even more clear cut.  Generic geodesics will remain
gravitationally bound to the halo, but will be spread chaotically
through micro-lensing.  (This means that the same event will lie on
multiple geodesics, but as discussed above, we will count each event
only once.)  Moreover, most large halos do not form directly from from
a single overdensity, but by mergers and accretion of smaller halos
that virialized much earlier.  One expects that most geodesics
threading merging halos will remain gravitationally bound during the
merger, and end up covering the resulting larger galaxy.  But then
observers will have a local scale factor which is less than the
averaged scale factor at the time of the formation of the smallest
structures that eventually merge to form our galaxies.  Note that
these structures need not even be galaxies, i.e., their mass could be
below $10^7$ solar masses.

The measure defined by De Simone {\em et al.} instructs us to include
all observations reached by geodesics whose maximum scale factor time
is below the cutoff.  Therefore, all observations in collapsed regions
will contribute as soon as the cutoff exceeds the maximum local scale
factor of the first collapsing objects that end up constituting the
host objects of observations.  

To include this effect, one could generalize to a more detailed
metric.  But it is much simpler to continue working with the
homogeneous FRW metric, Eq.~(\ref{eq-FRW}), and to include collapse
effects by pretending that all observations in collapsed regions
happen at the time $\tau^{\rm NL}$ when those regions decoupled from
the Hubble flow.  This amounts to projecting observations from the
time $\tau^{\rm obs}$ back to the earlier time $\tau^{\rm NL}$, and
thus, to increasing their number density per physical volume to
\begin{equation}
  \hat\rho_i^{\rm obs}\equiv 
  \rho_i^{\rm obs}\left(\frac{a_i^{\rm obs}}{a_i^{\rm NL}}\right)^3~,
\label{eq-rhohat}
\end{equation}
where $a_i^{\rm NL}\equiv a(\tau_i^{\rm NL})$.

Thus, we can include the effects of local collapse by replacing
Eq.~(\ref{eq-main}) by
\begin{equation}
  {\cal P}_i\propto e_i\, \hat\rho_i^{\rm obs} ~.
\label{eq-maini}
\end{equation}

\section{The cosmological constant}
\label{sec-cc}

In this section, we estimate the probability distribution for the
cosmological constant in the scale factor measure.  We will find that
it agrees roughly with the observed value in the no-collapse
approximation.  Treating collapsed regions according to the
prescription of De Simone {\em et al.}, however, yields a much larger
value.  We will then consider variations of the scale factor measure
and discuss how they might solve this problem.  Let us begin by
reviewing the history of anthropic predictions for the cosmological
constant and how they depend on the measure.

Throughout this paper, we will focus on positive values of $\Lambda$,
which have a greater potential for discrepancy with observation.
Negative values of $\Lambda$ are bounded (in any non-pathological
measure; for example, if there is no youngness paradox) by the order
of magnitude of the observed value, but positive values could be much
larger~\cite{Wei87}.  Both the causal diamond measure~\cite{BouHar07}
and the no-collapse scale factor measure~\cite{DGSV08} assign somewhat
higher integrated probability to negative values than to positive
values of $\Lambda$, but the imbalance is not large enough to render
the observed value unlikely.

\subsection{$\Lambda$ and observers per baryon}
\label{sec-opb}

In anthropic approaches to the cosmological constant problem, one
computes a probability distribution for {\em observed\/} values of the
cosmological constant.
Originally~\cite{Efs95,Vil95,Wei96,MarSha97,Vil04,GarSch05,Wei05} this 
was implemented by multiplying the underlying distribution (arguably, 
$dp/d\Lambda=$ const in the regime of interest) by the ``number of 
observers per baryon'', or per some other reference object.  
Implicitly, this defines an (incomplete) measure. If one tries to take 
``observers per baryon'' seriously as a measure, one finds that any 
eternally inflating de~Sitter vacuum has an infinite number of 
observers per baryon (if it has any observers at all), and 100\% of 
the observers are Boltzmann brains~\cite{Pag06,BouFre06b}.

Even aside from this problem, however, the observer-per-baryon measure
was already plagued by some annoying problems.  The simplest test of
the landscape solution to the cosmological constant problem is to
restrict attention to vacua that differ from ours {\em only}\/ through
their value of $\Lambda$.  In this setting, the observer-per-baryon
measure prefers values of $\Lambda$ about 5000 times larger than the
observed value; the observed value is excluded at the $3.5\sigma$
level~\cite{BouHar07}.  We find it useful to explain this in terms of
the timescale
\begin{equation}
\tau_\Lambda= \sqrt{\frac{3}{\Lambda}}
\end{equation} 
at which the cosmological constant comes to dominate the evolution of
the universe.  In the observer-per-baryon measure, the number of
observers will be proportional to the number of baryons that are
captured by galaxies.  The underlying distribution prefers larger
values of $\Lambda$.  There is no penalty until $\Lambda$ becomes
large enough to disrupt galaxy formation ($\tau_\Lambda \ll \tau_{\rm
  gal}$); such values will be strongly suppressed.  Thus one would
expect $\Lambda$ to be correlated with the time of the formation of
the first galaxies, $\tau_{\rm gal}$:
\begin{equation}
\Lambda\sim \tau_{\rm gal} ^{-2}~.  
\label{eq-lamgal}
\end{equation}

In fact, however, the observed value of $\Lambda$ appears to be
correlated with the (considerably later) time when observations are
made, $\tau_{\rm obs} $:
\begin{equation}
  \Lambda\sim \tau_{\rm obs} ^{-2}\sim 10^{-4} \tau_{\rm gal} ^{-2}~.
\label{eq-reallambda}
\end{equation}
This is the coincidence problem.  The observers-per-baryon measure
(ignoring Boltzmann brains) would have naturally explained a
hypothetical coincidence between $\tau_{\rm gal}$ and $\tau_\Lambda $, but
it is only barely consistent with the actual coincidence between
$\tau_{\rm obs}$ and $\tau_\Lambda $.

Further tests can be made by allowing other parameters (such as the
density contrast, curvature, etc.) to vary in addition to $\Lambda$,
or by integrating out such parameters altogether.  This exacerbates
the troubles of the observer-per-baryon measure.  For example, it
strongly prefers larger values of the density contrast, $Q$, than the
observed value $Q\sim 2\times 10^{-5}$.  Since $\tau_{\rm gal}\propto
Q^{-3/2}$, vacua with larger $Q$ can tolerate larger $\Lambda$ (like
$Q^3$) while still forming structure.  Thus, a greater fraction of
such vacua contains observers, and they are strongly preferred.

\subsection{$\Lambda$ and the causal diamond measure}
\label{sec-cd}

The causal diamond measure~\cite{Bou06,BouFre06a} was the first to
solve these problems.  Most directly, it solves the coincidence
problem, predicting a value of $\Lambda$ such that $\tau_\Lambda\sim
\tau_{\rm obs}$, or $\Lambda\sim \tau_{\rm obs}^{-2}$.  Greater values of
$\Lambda$ are suppressed mainly because observers become exponentially
dilute within the cutoff~\cite{BouHar07}:
\begin{equation}
  \frac{d{\cal P}}{d\log\Lambda}\propto\Lambda^{1/2}
  \exp\left(-\frac{3 \tau_{\rm obs}}{\tau_\Lambda}\right)~.
\label{eq-cd}
\end{equation}
Using $\tau_{\rm obs}\sim O(10$ Gyr$)$ (or alternatively, and less
anthropocentrically, equating the peak observation time with the time
of peak entropy production), this relation leads to a prediction of
$\Lambda$ in excellent agreement with the observed value.  Because the
causal diamond measure correlates $\Lambda$ directly with the era when
observations are made, and not with the time when structure forms, the
$Q$ runaway problem is absent~\cite{BouHar07,CliFre07}.

\subsection{$\Lambda$ and the scale factor measure}
\label{sec-sf}

De Simone {\em et al.}~\cite{DGSV08} have argued that the scale factor
measure shares some of these desirable properties.  We will now
examine this claim, focusing, once more, on the case where only
$\Lambda$ varies, while keeping $\tau_{\rm obs}$ fixed. We will find
that the conclusions of De Simone {\em et al.}  apply in the
no-collapse approximation, but are invalidated by collapse effects.

\subsubsection{No-collapse approximation}
\label{sec-noc}

In the no-collapse approximation adopted in Ref.~\cite{DGSV08},
Eq.~(\ref{eq-main}) tells us that
\begin{equation}
  \frac{d{\cal P}}{d\Lambda}\propto 
  \frac{dp}{d\Lambda} \rho^{\rm obs}(\Lambda)
\end{equation}
We assume that $dp/d\Lambda$ is constant in a small interval (say,
$10^{-20}$) around $\Lambda=0$, after coarse graining over an even
smaller interval (say, $10^{-130}$).  If this is not the case, then
the landscape cannot solve the cosmological constant problem, and the
scale factor measure would be ruled out.  Hence we may drop this
factor and write
\begin{equation}
  \frac{d{\cal P}}{d\log\Lambda}\propto 
  \Lambda\, \rho^{\rm obs}(\Lambda)~.
\label{eq-cc1}
\end{equation}

To understand how the observer density depends on $\Lambda$, let us
rewrite $\rho^{\rm obs}$ as a fraction of the physical matter density
at the time of observations,
\begin{equation}
  \rho^{\rm obs}(\Lambda) = 
  \alpha(\Lambda)\, \rho_{\rm matter}^{\rm obs}(\Lambda)~,
\label{eq-alphadef}
\end{equation}
so that 
\begin{equation}
  \frac{d{\cal P}}{d\log\Lambda}\propto 
  \Lambda\, \alpha(\Lambda)\, \rho_{\rm matter}^{\rm obs}(\Lambda)~,
\label{eq-ccalpha1}
\end{equation}

It is reasonable to assume that $\alpha$, the number of observers per
unit matter mass, is proportional to the Press-Schechter mass
function~\cite{MarSha97}, the fraction of matter mass in collapsed
objects above a certain critical mass ($10^7$ solar masses for the
smallest galaxies; more if one makes the additional anthropic
assumption that observers require particularly large galaxies).  To
first approximation, $\alpha$ is unaffected by $\Lambda$ as long as
$\tau_\Lambda\gg \tau_{\rm gal} $. Larger values of $\Lambda$ will
noticaby affect galaxy formation, and for $\tau_\Lambda\ll \tau_{\rm
  gal}$, $\alpha$ rapidly vanishes.

The matter density at the time of observations is independent of
$\Lambda$ as long as $\tau_\Lambda\gg \tau_{\rm obs}$.  But if
$\Lambda$ comes to dominate earlier, then it will drive a period of
exponential expansion before observations are made, and the matter
density will be far lower.  Roughly, we can write
\begin{equation}
  \rho_{\rm matter}^{\rm obs}(\Lambda)\approx
  \rho_{\rm matter}^{\rm obs}(\Lambda=0)~ 
  \exp\left(-3\, \frac{\tau_{\rm obs}}{\tau_\Lambda}\right)~.
\end{equation}
For example, in vacua with $\tau_\Lambda \approx 2$ Gyr, the density of
galaxies at $\tau_{\rm obs} \approx 13.7$ Gyr will a factor of order
$10^{-9}$ times smaller than in vacua with $\tau_\Lambda\gg 13.7$ Gyr.
Thus, moderately large values of $\Lambda$ are totally suppressed,
even though they do not disrupt galaxy formation.  We conclude that in
the no-collapse approximation, the suppression of large $\Lambda$ is
dominated by the dilution of matter, and one obtains
\begin{equation}
  \frac{d{\cal P}}{d\log\Lambda}\propto \Lambda
  \exp\left(-3\, \frac{\tau_{\rm obs}}{\tau_\Lambda}\right)~.
\end{equation}

Comparison with Eq.~(\ref{eq-cd}) reveals that the exponential factor
is the same as that arising in the causal diamond measure.  In both
measures, this factor arises primarily from the effect of $\Lambda$ on
the matter density, rather than its effect on galaxy formation.  The
prefactors differ because the physical volume of the causal diamond
also depends on $\Lambda$, like $\Lambda^{-1/2}$ for large $\Lambda$.
(In both cases the prefactor contains a factor $\Lambda$ from the
Jacobian $d\log\Lambda/d\Lambda$.)  Both distributions peak near
$\Lambda\sim \tau_{\rm obs}^{-2}$, in excellent agreement with
observation, Eq.~(\ref{eq-reallambda}).

\subsubsection{Proper treatment of collapsed regions}
\label{sec-col}

With the proper treatment of collapsed regions, however, the result
changes drastically.  Eq.~(\ref{eq-main}) is replaced by
Eq.~(\ref{eq-maini}), and correspondingly, we must replace
Eq.~(\ref{eq-cc1}) by
\begin{equation}
  \frac{d{\cal P}}{d\log\Lambda}\propto 
  \Lambda \hat\rho^{\rm obs}(\Lambda)~,
\label{eq-cc2}
\end{equation}
where $\hat\rho^{\rm obs}$ is the density, per physical volume, that
observations would have had if they had occurred at the time $\tau_{\rm
  NL}$ when the first dark matter halos decoupled from the Hubble
flow.  By Eqs.~(\ref{eq-rhohat}) and (\ref{eq-alphadef}), it follows
that
\begin{equation}
  \frac{d{\cal P}}{d\log\Lambda}\propto 
  \Lambda\, \alpha(\Lambda)\, \rho_{\rm matter}^{\rm NL}(\Lambda)~,
\label{eq-ccalpha2}
\end{equation}
where $\rho_{\rm matter}^{\rm NL}$ is the physical matter density at
the time $\tau_{\rm NL}$.

This matter density is insensitive to $\Lambda$ unless $\Lambda$ is
large enough to affect the formation of the earliest dark matter
haloes, at the time $\tau_{\rm NL}$.  Since $\tau_{\rm NL}<\tau_{\rm
  gal}$, the dilution of matter density is now no longer the effect
suppressing large $\Lambda$.  Rather, it is disruption of galaxy
formation: the Press-Schechter factor, which enters through $\alpha$.
It becomes suppressed if $\Lambda$ exceeds $\tau_{\rm gal}^{-2}$, so
the distribution will peak near this value.

Thus, the scale factor measure reproduces the result obtained from the
observers-per-baryon measure, Eq.~(\ref{eq-lamgal}), except for the
latter measure's manifest Boltzmann brain problem.  It prefers a value
of $\Lambda$ that is, depending on the strength of anthropic
assumptions, one to four orders of magnitude larger than the observed
value.  

Moreover, the scale factor measure suffers from a runaway problem if
the strength of initial density perturbations, $Q$, is allowed to
vary.  The pressure towards large $Q$ is actually greater than in the
observers-per-baryon measure.  In both measures, larger $Q$ allows for
larger values of $\Lambda$, since the maximum vacuum energy is of
order the density at the time of galaxy formation, which is
proportional to $Q^3$.  This enters through the first factor in
Eq.~(\ref{eq-ccalpha2}), suppressing the probability of the observed
value of $\Lambda$ by an additional factor $(Q_0/Q)^3$.  But in the
scale factor measure, the last factor in Eq.~(\ref{eq-ccalpha2})
offers an additional reward for large $Q$: the density at the time of
the earliest halo formation also scales like the third power of $Q$.
This makes it more difficult to compensate for the runaway problem by
prior distributions or anthropic cutoffs.

\subsection{Can the scale factor measure be improved?}
\label{sec-improv}

We have found that the scale factor measure predicts $\Lambda\sim
\tau_{\rm gal}^{-2}$, while the causal diamond measure predicts
$\Lambda\sim \tau_{\rm obs}^{-2}$.  Can we pinpoint the key difference
between the measures leading to these different results?  And can the
definition of the scale factor measure be modified to yield a more
desirable answer?

The causal diamond measure is defined nonlocally, in terms of the
event horizon of a single worldline.  Observations inside this horizon
will be included, those outside will not.  The horizon size is not
affected by local phenomena such as gravitational collapse.  If
$\tau_\Lambda\ll \tau_{\rm obs}$, then observations will be dilute, and few
if any will be captured by the diamond.

The scale factor measure is defined in terms of a local quantity, the
expansion and local scale factor of a congruence of geodesics.  The
local scale factor runs backwards once structure formation begins, and
a special rule is needed to decide how the cutoff should be specified
in this case.  By the rule of De Simone {\em et al.}, future
observations occurring on such geodesics are immediately counted at
structure formation, so the dilution of galaxies by the cosmological
constant cannot affect the probability distribution.

One might attempt to improve the scale factor measure by replacing the
local scale factor, $A$, with the averaged scale factor, $a$,
describing the spatial curvature radius in expanding open FRW bubbles,
while a geodesic is passing through matter dominated regions (up to an
obvious constant factor that depends on the scale factor time at which
the geodesic enters the matter dominated region).  The idea would be
to avoid the effects of collapse by averaging over collapsed regions
and defining the scale factor time in terms of the averaged Hubble
flow.

A problem with this approach is that the FRW scale factor $a$ is only
approximately defined.  On the scales currently within the horizon,
there exists a preferred slicing of constant average density, which
defines hypersurfaces of constant $a$.  However, because the universe
is not exactly homogeneous and isotropic, the averaging procedure is
necessarily somewhat ambiguous---in the observable universe, at least
at the level of $10^{-5}$.  At larger scales, or in different bubbles,
density fluctuations can be much larger.  Bubbles with eternal
slow-roll inflation cannot be assigned a preferred FRW slicing at all;
neither can the asymptotic regions containing Boltzmann
brains.  Thus, it is unclear how this idea would lead to a
sufficiently general prescription for computing probabilities.


A more promising approach to improving the scale factor measure is to
change the rule that deals with collapsing geodesics, so that
observers are not counted by geodesics that turn around too early.
For example, whenever an event is passed through by multiple
geodesics, we might assign its scale factor time to be the {\em
  maximum} scale factor time among those geodesics.  This is in
principle as well-defined as the proposal in \cite{DGSV08}, which
corresponds to using the minimum.  A closely related idea is to extend
each geodesic only until its first caustic, when neighboring geodesics
intersect and $\theta\to-\infty$.  Whether such prescriptions lead to
a better prediction for $\Lambda$ is an interesting question, but one
beyond the scope of the present paper.

Finally, we may turn to the local formulation of the scale factor
measure, Eq.~(\ref{eq-fatline}), for help.  In expanding regions, it
is equivalent to the formulation using a geodesic congruence, but as
noted at the end of Sec.~\ref{sec-local}, the local formulation can be
applied without modification in collapsed regions.  It would, however,
suffer from a similar problem with predicting $\Lambda$.  Most ``fat
geodesics'' are captured by dark matter halos around the time when
those objects first form.  Therefore, the expected density of
observations is set by the matter density inside galaxies, and not by
the large scale average of the matter density of the FRW solution at
the time of observations, $\tau_{\rm obs}$.  As a result,
Eq.~(\ref{eq-fatline}) is insensitive to the value of the cosmological
constant unless it is large enough to disrupt galaxy formation.  It
would predict $\Lambda\sim \tau_{\rm gal}^{-2}$, in poor agreement
with the observed value $\Lambda\sim \tau_{\rm obs}^{-2}$.

Still, Eq.~(\ref{eq-fatline}) may turn out a useful starting point for
a more successful measure.  Instead of fattening the geodesic
infinitesimally, for example, we could consider constructing a larger
transverse volume.  We can then either attempt to take a large volume
limit, in which the average density may become sensitive to the
dilution caused by early vacuum domination.  (Whether this can be done
in a well-defined manner is, again, a question beyond the scope of our
present work.)  Or we could use a finite transverse volume.  To avoid
explicitly introducing an arbitrary length scale into the measure,
this volume could be defined in terms of the cosmological event
horizon surrounding the worldline.  This fixes the $\Lambda$ problem,
but it fails to produce a novel measure: With this prescription, we
would simply recover the causal-diamond measure (with a particular
choice of initial conditions).

\section{Boltzmann brains}
\label{sec-BB}

In this section, we compute the number of Boltzmann brains, and
determine the conditions under which they dominate over ordinary
observers.  We define Boltzmann brains
as observers that arise from local violations of the second law of
thermodynamics.  This occurs at late times in de~Sitter vacua.  States
of energy $E\gg T$ are produced at the Boltzmann-suppressed rate
\begin{equation}
\Gamma_i(E) = {4\pi \over 3} H_i^{4}
\exp(-E/T_i)
\end{equation}
where $T_i=H_i/2\pi$ is the Gibbons-Hawking temperature of the
de~Sitter horizon.  The minimum mass of a Boltzmann brain (if any)
will depend on the vacuum.  But on general
grounds~\cite{Bou00b,BouFre06b}, the exponential factor cannot be
larger (though it can be much smaller) than $\exp(-S_{\rm BB})$, where
$S_{\rm BB}$ is the course-grained entropy, or number of particles, in
the most primitive Boltzmann brain,
\begin{equation}
\Gamma_i^{\rm BB} < \exp(-S_{\rm BB})~.
\label{eq-bbd}
\end{equation}
One can only speculate about the value of $S_{\rm BB}$, but it is
certain to be exponentially large.  Thus, Boltzmann brains are
double-exponentially suppressed.

\subsection{The ratio of Boltzmann brains to ordinary observers}
\label{sec-ratio}

We can neglect crunching vacua, as well as the initial non-de~Sitter
regime of metastable vacua: because of the enormous suppression of
Boltzmann brains, almost all of them will be produced in the
asymptotic de~Sitter regime.
The number of Boltzmann brains produced in vacuum $i$ between the time
$\eta$ and $\eta+d\eta$ is
\begin{equation}
  dN_i^{\rm BB}=\Gamma^{\rm BB}_i V_i d\eta/H_i~.
\end{equation}
By Eq.~(\ref{eq-volumes}), the total number of Boltzmann brains
produced prior to the cutoff $\eta$ in vacuum $i$ is therefore
\begin{equation}
  N_i^{\rm BB}=\frac{Ce^{\gamma\eta}}{\gamma}
  \frac{s_i}{H_i} \Gamma^{\rm BB}_i ~.
\end{equation}
By Eqs.~(\ref{eq-sstar}) and (\ref{eq-eigen}), $s_*\approx 1$ for the
dominant vacuum and $s_i=p_i/(\kappa_i-q) \approx p_i/\kappa_i$ for
all other vacua.  Dropping the usual factor $Ce^{\gamma\eta}/\gamma$,
the total probability for observations by Boltzmann brains is
\begin{equation}
  {\cal P}^{\rm BB} \propto
  H_*^{-1} \Gamma^{\rm BB}_* +
  \sum_{i\neq *} H_i^{-1} p_i\frac{\Gamma^{\rm BB}_i}{\kappa_i} ~.
\end{equation}
We are interested in comparing this to the total number of
observations by ordinary observers,
\begin{equation}
  {\cal P}^{\rm OO}_i \propto \sum_i p_i \rho^{\rm OO}_i
\end{equation}

The key simplification arising in this comparison is that $p_i$,
$\Gamma^{\rm BB}_i$, and $\Gamma_i$, are generically double
exponentials.  That is, they are of the form $\exp(\pm \exp x)$ with
$x\gg 1$.  We will use a triple inequality sign for such numbers, for
example
\begin{equation}
\Gamma^{\rm BB} \lll 1
\end{equation}
Such numbers obey special laws of arithmetic.  For example, for $y$
and $z$ double-exponentially large, $y/z\approx y$ if $y>z$.
Moreover, if $y$ is a single exponential and $z$ a double exponential,
then $zy\approx z/y\approx z$.  A double exponential takes the same
value in any conventional system of units, though it can be useful to
think in terms of Planck units for definiteness.

The landscape contains an exponentially large number of vacua, but so
far there are no indications that it might be doubly-exponentially
large.  Thus, $H_i$ is at most a single exponential, and we can write
\begin{equation}
  \frac{{\cal P}^{\rm BB}}{{\cal P}^{\rm OO}}\approx \frac{
    \Gamma_*^{\rm BB}+\sum_{i\neq *} p_i~(\Gamma_i^{\rm BB}/\Gamma_i)}{
    \sum_i p_i\,\rho^{\rm OO}_i}
\label{eq-ratio}
\end{equation}
Note that we have retained $ \rho_i^{\rm OO}$, since it can be zero or
doubly-exponentially small in some vacua.  Because
Eq.~(\ref{eq-ratio}) involves a ratio of double exponentials, either
the numerator or the denominator will completely dominate (depending
on the measure, and on the landscape), and the relative probability
will be zero or infinity to good approximation.  Whoever wins, wins
big.

We can restate this result in terms of the expected number $e_i$ of
times a worldline starting in the $*$ vacuum will pass through vacuum
$i$ (see the discussion at the end of Sec.~\ref{sec-defrate}).  Using
$q\approx\kappa_*$, we can include the $*$ vacuum in the sum:
\begin{equation}
  \frac{{\cal P}^{\rm BB}}{{\cal P}^{\rm OO}}\approx \frac{
    \sum_i e_i~(\Gamma_i^{\rm BB}/\Gamma_i)}{
    \sum_i e_i\,\rho^{\rm OO}_i}~.
\label{eq-ratio2}
\end{equation}

At this level of approximation, the causal diamond measure yields
nearly the same result~\cite{BouFre06b}. In the causal diamond measure,
the number of Boltzmann brains in vacuum $i$ is the expected number of
times the generating worldline enters vacuum $i$, $e_i$, times
$\Gamma^{\rm BB}_i$, times the expected four-volume the diamond will
span in vacuum $i$ (given by the life-time of vacuum $i$,
$H_i^{-1}\kappa_i^{-1}$, times the de~Sitter horizon volume,
$4\pi/3H_i^3$).  The expected number of ordinary observers is $e_i$
times the number of ordinary observers inside the de~Sitter horizon of
vacuum $i$.  Keeping only double-exponentials, the relative
probability in the causal diamond measure is again given by
Eq.~(\ref{eq-ratio2}).

Aside from negligible non-double-exponential factors, the only way the
two measures differ, for the purposes of Boltzmann brains, is through
$e_i$, which depend on initial conditions.  A priori, the question of
initial conditions has nothing to do with the measure problem, and in
the causal diamond measure, they remain separate issues. One imagines
that the universe started, perhaps, in a randomly chosen vacuum, or in
an ensemble governed by the tunneling wavefunction, favoring
Planck-scale vacua.  Physical probabilities are not strongly affected
by this uncertainty~\cite{BouYan07}.  In the scale factor measure, on
the other hand, the eternally inflating universe exhibits attractor
behavior, which can be mimicked by choosing the particular initial
probability distribution $s_i$, defined in Sec.~\ref{sec-defrate}.
Roughly, this means starting the worldline in a very particular
vacuum: the longest-lived metastable vacuum~\cite{SchVil06}, $*$.

\subsection{Under what conditions do Boltzmann brains dominate?}
\label{sec-conditions}

In this subsection, we will show that with some reasonable assumptions
about the landscape, the ratio of Boltzmann brains to ordinary
observers takes an even simpler form.  Except for the $*$ term, each
term in the numerator of Eq.~(\ref{eq-ratio}) contains the ratio
$\Gamma_i^{\rm BB} / \Gamma_i$. By the laws of double exponential
arithmetic, this ratio is dominated by the larger of the two
exponents,
\begin{equation} 
{\Gamma_i^{\rm BB} \over \Gamma_i} \approx \left\{\begin{array}{rcl}
\Gamma_i^{\rm BB} 
&<& \exp(- S_{\rm BB}) \ \ {\rm if} \ \Gamma_i^{\rm BB} < \Gamma_i \\
\Gamma_i^{-1} \  
&>& \exp(+ S_{\rm BB}) \ \ {\rm if}  \
\Gamma_i^{\rm BB} > \Gamma_i\end{array}\right. ~,
\label{eq-jkl}
\end{equation} 
where we have used Eq.~(\ref{eq-bbd}) for the inequalities.

We will now constrain the factor $p_i$ multiplying this ratio.  With
no loss of generality, we may restrict the sum in Eq.~(\ref{eq-ratio})
to the set of vacua containing observers or Boltzmann brains.  Let us
label the vacua in this set by the index $a$ instead of $i$.  By
Eqs.~(\ref{eq-panth}) and (\ref{eq-piei}), $p_a$ can be obtained by a
sum, over all decay paths, of products of branching ratios.
\begin{equation} 
  p_a = q
  \sum_{{\rm paths~from}~*~{\rm to}~a} \beta_{a i_n}~
  \beta_{i_n i_{n-1}}~ ... ~\beta_{i_1 *}~~~(a\neq *)~,
\label{eq-panthstar}
\end{equation}
This equation holds only for $a\neq *$, but those are precisely the
$p_a$ appearing in Eq.~(\ref{eq-ratio}).\footnote{While our discussion
  will focus on the scale factor measure, it can be adapted to the
  causal diamond measure simply by replacing $*$ with another initial
  vacuum.}

We will make three assumptions:
\begin{itemize}
\item[(1)]{For every vacuum $a$, every decay path from $*$ to $a$ contains
    at least one intermediate de~Sitter vacuum.---This assumption is
    very plausible in a multidimensional landscape such as that of
    string theory, where neighboring vacua have vastly different
    values of the cosmological constant (a crucial feature for solving
    the cosmological constant problem~\cite{BP,TASI07}).  Assuming that
    distant vacua cannot be accessed with appreciable probability, it
    follows that $\Delta\Lambda\gg 1/S_{\rm BB}$ in direct decays, so
    it is exponentially unlikely that the dominant vacuum can decay to
    any vacuum large enough to contain observers or Boltzmann brains.}
\item[(2)]{For every vacuum $a$, there exists at least one decay path
    from $*$ to $a$ which does not pass through any de~Sitter vacua
    with horizon entropy bigger than $S_{\rm BB}$.---This assumption,
    too, is plausible given the scarcity of vacua with $\Lambda\ll 1$,
    and given our ability to choose any path we like.}
\item[(3)]{Of the decay paths posited in assumption (2), at least one
    ``dominates'' the sum in Eq.~(\ref{eq-panthstar}), in the
    following extremely weak sense: By dropping all other terms, we
    change the sum by less than a double exponential factor.---Like
    assumption (1), this is plausible in a multidimensional landscape
    with large step size, where decay chains in the semi-classical
    regime are short.}
\end{itemize}
We remain agnostic about whether the $*$ vacuum has horizon entropy
larger or smaller than $S_{\rm BB}$.  The dominant vacuum is special.
It is at least conceivable that its defining properties strongly
select for a very small cosmological constant, and we do not wish to
prejudice this issue here.

We will be interested in bounding $p_a$ only to within
double-exponentially large factors.  By assumption (3), we can drop
the sum over paths and write
\begin{equation} 
  p_a \sim q~
   \beta_{a i_n}~
  \beta_{i_n i_{n-1}}~ ... ~\beta_{i_1 *}~~~(a\neq *)~,
\label{eq-nosum}
\end{equation}

Assumption (1) guarantees that that $n\geq 1$, so $\beta_{i_1 *}$
exists in the product and represents the branching ratio from one
de~Sitter vacuum to another.  For two connected de~Sitter vacua,
detailed balance gives $\kappa_{i_1*} = \kappa_{*i_1} \exp(S_{i_1} -
S_*)$ where $S_j$ is the horizon entropy of vacuum $j$. Also, the
decay rate must be faster than the recurrence time, $\kappa_{i_1*} >
\exp(-S_*)$. We are only considering vacua which eternally inflate,
$\kappa_{*i_1} < 1$, so
\begin{equation}
\exp(-S_*) < \kappa_{i_1*} < \exp(S_{i_1} - S_*)~.
\end{equation}
We can divide by $\kappa_*$ to get a bound on the branching ratio,
\begin{equation}
{1 \over \kappa_*} \exp(- S_*) < 
\beta_{i_1*} < {1 \over \kappa_*} \exp(- S_* + S_{i_1})~.
\label{eq-dombound}
\end{equation}
For the other branching ratio(s) appearing in Eq.~(\ref{eq-nosum}), we
can use the cruder bound
\begin{equation}
\exp(- S_j)
 < {1 \over \kappa_j} \exp(-S_j)
 <  \beta_{ij} < 1~,
\end{equation}
which holds also if the destination vacuum is terminal (as $a$ might
be).  Substituting these bounds into Eq.~(\ref{eq-panth}), we obtain
\begin{equation}
  \exp(- S_*) \exp(- S_{i_1} - S_{i_2} - ... - S_{i_n}) <   p_a
  <   \exp(- S_* + S_{i_1})~,
\end{equation}
where we have used $\kappa_* \approx q$.  By assumption (2) (and
assuming that the path is not exponentially long),  we
have $\sum_{k=1}^n S_{i_k} \ll S_{\rm BB}$, so
\begin{equation}
  \exp(- S_*) \exp(- S_{\rm BB}) \lll p_a 
  \lll  \exp(- S_*) \exp(+ S_{\rm BB})~.
\end{equation}
(We remind the reader that we use the triple inequality sign for the
inequality of double exponentials.)

We have bounded $p_a$ to within a range small compared to $\exp(S_{\rm
  BB})$.  Meanwhile, by Eq.~(\ref{eq-jkl}), the ratio $\Gamma_i^{\rm
  BB}/ \Gamma_i$ is either larger than $\exp(S_{\rm BB})$ or smaller
than $\exp(-S_{\rm BB})$.  By double exponential arithmetic,
therefore, we can neglect the uncertainty in $p_a$ when we multiply
these terms:
\begin{equation}
  p_a {\Gamma_a^{\rm BB} \over \Gamma_a}  
  \approx  \exp(- S_*) {\Gamma_a^{\rm BB} \over \Gamma_a} 
\end{equation}
By similar double exponential reasoning we can do the same thing in
the denominator, so the ratio becomes
\begin{equation}
{{\cal P}^{\rm BB} \over {\cal P}^{\rm OO}} = 
{\Gamma_*^{\rm BB} + \sum_a \exp(- S_*) (\Gamma_a^{\rm BB} /\Gamma_a) 
\over \sum_a \exp(-S_*) \rho_a^{\rm OO}}
\end{equation}
It seems likely that in our vacuum the density of observers is not
double exponentially small, and in Planck units $\rho^{\rm OO} < 1$,
so $\sum_a \rho_a^{\rm OO}$ is not double exponential as long as the
number of vacua is not double exponential. Therefore,
\begin{equation}
{{\cal P}^{\rm BB} \over {\cal P}^{\rm OO}} =  
\exp(S_*) \Gamma_*^{\rm BB} + \sum_a {\Gamma_a^{\rm BB} 
\over \Gamma_a} 
\end{equation}
For the ordinary observers to dominate, the ratio must be less than one, so every term must be less than one. This requires
\begin{equation}
  {\Gamma_a^{\rm BB} 
    \over \Gamma_a} < 1 \ \ \forall~ a~.
\label{eq-con1}
\end{equation}
Also, if the dominant vacuum can produce Boltzmann brains at all, then
it will do so far faster than the recurrence rate $\exp(-S_*)$, so for
the first term to be less than one we need
\begin{equation}
\Gamma_*^{\rm BB} = 0~.
\label{eq-con2}
\end{equation}
If either of these conditions is violated, the Boltzmann brains
dominate.~\footnote{Our result is consistent with the conditions found
  by Linde~\cite{Lin06} for a toy landscape with two de~Sitter vacua
  and a sink.  This toy landscape is so small that it violates our
  assumptions, but nevertheless Eqs.~(\ref{eq-con1}) and
  (\ref{eq-con2}) are necessary and sufficient for ordinary observers
  to dominate.  The condition ``$\Gamma_{1s}\gg\Gamma_{21}$'' (in the
  notation of Ref.~\cite{Lin06}) ensures that $2=*$; the condition
  ``$\Gamma_{1s}\gg\Gamma_{1B}$'' corresponds to Eq.~(\ref{eq-con1}).
  Eq.~(\ref{eq-con2}) was not explicitly spelled out but is implicit
  in Fig.~4 of Ref.~\cite{Lin06}.}

\subsection{Discussion}

Let us summarize our result in general terms.  If any vacua have
lifetimes longer than their Boltzmann brain time
($\Gamma_a<\Gamma_a^{\rm BB}$), then by the laws of double
exponentials, the Boltzmann brains will dominate---unless there is a
conspiracy in the landscape so that the rate of production $p_a$ of
every single Boltzmann brain producing vacuum is double exponentially
small compared to the $p_a$ for some vacuum which contains mostly
ordinary observers. Conversely, if all vacua decay before they produce
Boltzmann brains, and the $*$ vacuum does not produce Boltzmann
brains, then ordinary observers dominate unless all of the $p_a$ for
ordinary observer vacua are double exponentially small compared to the
$p_a$ for some vacuum which produces primarily Boltzmann brains.

It is interesting to compare the conditions necessary for the absence
of Boltzmann brains with those arising in the causal diamond measure.
For the purposes of Boltzmann brains, the causal diamond measure and
the scale factor measure differ only through the choice of initial
conditions, i.e., through the second of the two conditions in
Eqs.~(\ref{eq-con1}) and (\ref{eq-con2}).  In the causal diamond
measure, there is no reason to select initial conditions that favor
vacua with extremely small cosmological constant $\Lambda<1/S_{\rm
  BB}$.  Thus, it is implausible that Boltzmann brains would dominate
via Eq.~(\ref{eq-con2}) (with the $*$ vacuum replaced by the relevant
initial conditions).  

In the scale factor measure, on the other hand, the $*$ vacuum may
have very small cosmological constant, for example because of a
correlation between vacuum stability and the degree of supersymmetry
breaking.  Thus, while the string landscape may well satisfy
Eq.~(\ref{eq-con1}), it appears that the scale factor measure gives
Boltzmann brains a second chance through Eq.~(\ref{eq-con2}).

We should not make too much of this difference.  Cosmological constant
aside, Boltzmann brains are fairly complex objects and will not arise
in every imaginable low-energy field theory.  Thus, we expect
$\Gamma_a^{\rm BB}$ to vanish exactly in most vacua.  In the $*$
vacuum, the necessary conditions on matter content are no more likely
to be satisfied than in any other randomly chosen vacuum, since
low-energy properties like particles and fields are unlikely to be
correlated with the high-energy features responsible for the vacuum's
longevity.  Thus it seems very likely that $\Gamma^{\rm BB}_*=0$.

\acknowledgments We thank A.~Guth and A.~Vilenkin for discussions.
This work was supported by the Berkeley Center for Theoretical
Physics, by a CAREER grant (award number 0349351) of the National
Science Foundation, and by the US Department of Energy under Contract
DE-AC02-05CH11231.

\bibliographystyle{board}
\bibliography{all}

\end{document}